# Sorption heat engines: simple inanimate negative entropy generators

Anthonie W. J. Muller and Dirk Schulze-Makuch

Corresponding author: Anthonie Muller
e-mail: awjmuller@wsu.edu, Tel: 509-335-1501, fax: 509-335-7816

AFFILIATION:
Department of Geology, Washington State University, Pullman WA 99164-2812, USA

**Abstract**
The name 'sorption heat engines' is proposed for simple negative entropy generators that are driven by thermal cycling and work on alternating adsorption and desorption. These generators are in general not explicitly recognized as heat engines. Their mechanism is applicable to the fields of engineering, physics, chemistry, geology, and biology, in particular the origin of life. Four kinds of sorption heat engines are distinguished depending on the occurrence of changes in the adsorbent or adsorbate during the thermal cycle.

## 1. Introduction

The proposed sorption heat engine is based on adsorption and desorption, which are of large practical importance [1-6] but receive scant notice outside physics and engineering.
The engine works by desorption at high temperature, with concomitant uptake of heat, and release of the adsorbate at high pressure (Figs 1-2). After having performed work, the adsorbate is readsorbed at low pressure and low temperature with release of heat. The reverse engine has found application in refrigerators/heat pumps that have attracted much interest. The forward engine plays however only a minor role, as subsystem in complex refrigerators/heat pumps. This study points to the possible more general and extended role of the sorption heat engine in physics, chemistry, geology and biology, where its simplicity finds use in modeling the self-organization of inanimate matter that resulted in the origin of life:
1. In physics: in addition to the described refrigerators, the combination with convection could permit the gain of high free energy compounds from this self-organizing dissipative structure;
2. In chemistry the engine's relevance is not as a free energy source, but as a basic mechanism behind the local lowering of entropy by demixing. Adsorption followed by desorption is widely applied for separation/demixing and purification in analytical chemistry and chemical engineering. Another standard separation method is distillation, in which heat moves from high to low temperature as well, with an entropy lowering by demixing as the result. Since the distinction between a heat engine and a distiller becomes blurred in this context, we call a heat engine a device that increases the Gibbs free energy while heat moves from high to low temperature. The term 'negative entropy generator' is clearly applicable;
3. In geology, desorption of water and organics from clay minerals occurs on a large scale during subduction in the Earth's crust, a process driven by convection in the Earth's mantle;

4. In biology the engine may have played a role in the synthesis and desorption of the first biomolecules adsorbed on clay [7-9], and may thus have constituted an alternative source for the primordial soup proposed by Haldane in 1929 [10].

The concept yields a unifying point of view on phenomena hitherto considered subject of distinct scientific disciplines. It can be extended to include the proposed synthesis of compounds with a high Gibbs free energy of formation. This permits a classification in four kinds according to the constancy of adsorbate and adsorbens during the thermal cycle (Fig. 3):
1. Both the adsorbent and the adsorbate remain unchanged;
2. The adsorbent changes, but the adsorbate remains the same;
3. The adsorbent remains the same, but the adsorbate is transformed to a compound with a higher free energy of formation;
4. Both the adsorbent and the adsorbate change, the latter being transformed to a compound with a higher free energy of formation.
It is often advantageous to store the free energy produced by an engine [11]. Where a high free energy compound is generated—kinds 3 and 4—such storage is automatic.

## 2. Physics and Thermal Engineering

There is much more to adsorption than the weak bonding of an adsorbate molecule to the adsorbent's surface. Large physical changes can occur. The pressure can drop to very low values. The heat of adsorption of a vapor can be several times larger than its heat of condensation, and the combined volume of adsorbate and adsorbent can be less than the volume of the condensed liquid [1], which suggest a state of matter intermediate between the solid and the liquid state. The heat released during adsorption can be so large that safety precautions need to be taken [12].

The combination of adsorption with desorption is of interest as well. Temperature controlled adsorption-desorption cycles have been described and are widely applied [1, 6]. In the proposed 'sorption heat engine' (SHE), adsorption at low temperature with release of heat is followed by desorption at high temperature with uptake of heat (Figs 1 and 2). The $p$-$V$ diagram of the SHE differs from the $p$-$V$ diagrams of the Sterling engine and a vapor based engine such as the steam engine (Fig. 4). The most simple kind of heat engine, in which both the adsorbent and adsorbate do not change, is called a 'sorption heat engine of the first kind' (SHE1) to distinguish it from sorption heat engines in which such changes do occur.

Since adsorption is a low temperature phenomenon, the SHE can operate only over a small temperature range, which lowers the Carnot ratio and thus the efficiency of the heat to work conversion. The efficiency is further lowered due to the inherent waste of the heat involved in changing the temperature of the bulk adsorbent. Although adsorption and desorption often are reversible, they can be slow [1]. The SHE operates on a time scale larger than that of most machines. The low efficiency and power rate seem to make the application as free energy generation device not practical. Consisting of only two components, the SHE is however very simple. There is also a wide choice of adsorbents. Clay, zeolites, activated carbons, silica gel, and alumina (bauxite) are commonly used [1-6], and for special purposes specific minerals or materials may be more suitable.

Simple small refrigerators without moving parts have been described that use solar heat and operate by variable adsorption and desorption of a vapor (Fig. 5) [5, 13-15]. Engineering journals such as the *International Journal of Refrigeration* and *Applied Thermal Engineering* nowadays contain many studies of more complex adsorption-based refrigerators/heat pumps. In some studies the working cycles are analyzed in terms of Carnot cycles and subsystems that perform internally-used work [16-17]: see for instance Fig. 3 in ref. [17]. The overall goal in these studies is however still to effect cooling, and not to perform external work or demixing as proposed in this paper.

Adsorption is a rather general process not restricted to gases or vapors: larger molecules are adsorbed as well. Adsorption from the liquid phase is common but has not been studied as extensively [3].

The SHE requires thermal cycling, which can be effected by the engine being carried along by the current in a convection cell. Prigogine *et al.* [18] have emphasized the importance of self-organization in inanimate matter, such as for the origin of life, and introduced the concept of a non-equilibrium 'dissipative structure' that can perform external work. Because of the Second Law of Thermodynamics, such external work requires simultaneous dissipation. A major inanimate self-organizing dissipative structure in the world around us is the convection cell [18, 19], which is ubiquitous in the atmosphere's weather phenomena [20], in natural waters [21], and in the Earth's mantle [22]. The convection cell can perform external work, for instance in weather-driven windmills, but in natural systems the energy produced by external work is seldom stored (an example of storage in natural systems —as gravitational potential energy —is the deposition of snow and rain by the weather in snow fields and lakes at higher elevations). In general the available free energy is dissipated due to viscous effects in the convecting fluid. In a combination of convection with the SHE, the thermal cycling can drive the engine: thus a simple self-organized system is obtained that can yield external work and storable energy (Fig. 6).

## 3. Chemistry and Chemical Engineering

The SHE1 can be used for demixing [4]. Information gain by compound identification is a key object in analytical chemistry, often achieved by detection after demixing/separation such as by chromatography [23]. Here adsorption is followed by desorption, sometimes effected by a change in medium composition, sometimes by a change in temperature.

In gas chromatography (GC), for example, a sample containing several components is injected into a column where they are bound. Upon application of an increasing temperature, the desorbed components enter at different times the carrier gas flowing through the column, resulting in separation according to desorption temperature, and permitting component detection [24]. In liquid chromatography (LC), 'temperature-responsive chromatography' is an emerging technique [25, 26] that works on a phase transition of a polyacrylamide gel to a dehydrated state. Such a system concerns a 'sorption engine of the second kind' (SHE2).

Temperature programmed desorption in catalysis research [27] resembles the SHE1. In chemical engineering, separation and purification by variable adsorption are also widely applied: in the 'pressure-swing method' by varying the pressure, in the 'thermal swing method' by a thermal cycle [2-4].

Adsorption plays a key role in catalysis [28]. Materials such as clay minerals and zeolites are known for both their adsorptive and catalytic capabilities. When adsorbed, the molecules find themselves in a two dimensional environment, which increases the chance of collision, and thus the reaction rate. Adsorption can change a chemical equilibrium (demonstrated by pK changes of the adsorbate [29]), and the reaction rate may be further increased by a decrease of the activation energy of the reaction.

Chemical reactions decrease the free energy. When not coupled to another reaction, a single chemical reaction can only be exergonic [30]. After the exclusion of man-made engines, there are few sources for the free energy of formation of high energy compounds in the inanimate world. Any additional source is therefore of interest. One of the purposes of this paper is to show that the 'sorption heat engine of the third kind' (SHE3) could be a source of free energy, the free energy of formation.

Letting the reaction
$$R \to P,$$
go forward is impossible if the product P has a *higher* free energy of formation than the reactant R. The adsorbed product $P_{ads}$ could however spontaneously be formed from the adsorbed reactant $R_{ads}$, provided $P_{ads}$ has a *lower* free energy than $R_{ads}$. Obviously this $P_{ads}$ cannot be spontaneously released and must remain bound. Upon a rise of the temperature, the strength of adsorption decreases: at a high enough temperature, every species is desorbed [6]. By heating we can therefore obtain the searched-for high energy product P. In the following step we let P fall back to R, and use their difference in free energy to perform external work. The cooling of the adsorbent and R to their original temperature closes the cycle. The overall reaction is

$$\text{heat} \to \text{work},$$

with partial processes

$$R + \text{heat} \to P, \quad P \to \text{work} + R.$$

During the desorption of P the energetically allowed process

$$P_{ads} \to R_{ads} \to R,$$

must be kinetically suppressed. This suppression is similar to quenching,[1] the difference being its increase with the temperature, which is attributed to the diminished intensity of adsorption at higher temperature.

    The phenomenon of desorption upon a phase transition by the adsorbent in the SHE2 is generalized in the concept of the 'sorption heat engine of the fourth kind' (SHE4), in which the desorbed adsorbate also has a higher free energy of formation. In the SHE4 the just mentioned kinetic suppression may be effected by the change in the adsorbent.

    The reaction cycle of the SHE3 and SHE4 mechanisms is synchronized with the thermal cycle, which in general will be longer than the turnover time of regular catalysts.

    There is some direct experimental evidence of the conjectured SHE3. There are reports of substances changed upon thermally stimulated desorption [1]. The synthesis of high-energy compounds during adsorption is well known, such as the synthesis of small peptides from amino acids [31]. Although in the laboratory desorption of the high-energy compounds is commonly effected by an isothermal change in composition of the medium — for instance by the addition of a salt such as pyrophosphate [32] — there are also reports of desorption by thermal cycling, which was explicitly identified as the energy source [33].

## 4. Geology

The naturally occurring minerals clay and zeolite are adsorbates with a large, temperature-dependent adsorption [5,11-13]. When present in convection cells, particles of these minerals will be carried along, be thermally cycled and may function as SHE. Examples of convection cells in geology are the convection currents in volcanic hot springs and the convection currents in the Earth's mantle that drive plate tectonics [22].

    Many adsorbents such as silica and alumina are stable at temperatures of 200 C or higher. Other adsorbents are however unstable. The common clay montmorillonite undergoes the smectite to illite transition upon heating above a temperature of about 100 C (higher values up to 175 C have also been reported [34]). The transition causes desorption of adsorbed water [35-37]. In the Earth's crust the transition occurs simultaneously with the generation of oil [34-36, 38]; this 'crude oil' or 'petroleum' has its origin in organics adsorbed while the clay is near the Earth's surface [35].

    In the petrochemical industry adsorption by clay at temperatures below 100 C is used for the removal of oil components (mainly for discoloration and odor removal) [1]. By high temperature steam, the adsorbates are desorbed again and the adsorbent is regenerated, which demonstrates the applicability of a cyclic thermal process to the demixing of oil components.

---

[1] A temperature-dependent chemical equilibrium is called 'quenched' when cooling impedes the reaction of the reactants, making it impossible to attain the concentrations corresponding to the low-temperature equilibrium: the high-temperature equilibrium concentrations are frozen in, resulting in a disequilibrium that is a potential source of chemical free energy.

An example of the SHE2 is water adsorbed to clay and released by the smectite to illite transition. Desorbed oil components with a higher free energy of formation than the original reactants would be an example of an SHE4.

In Figure 7 we consider a possible application of these concepts. Convection in the Earth's mantle drives circulation in the Earth's crust. The clay cycle shown in Fig. 7 drives the cycling of adsorbates. Water-soluble organics desorbed from clay during the smectite to illite transition might not end up in the rather hydrophobic petroleum: instead, at a temperature above 100 C, these organics would diffuse quickly into the groundwater. When they have a high chemical free energy, the organics may constitute the food source for some of the microorganisms found in the Earth's crust [39-43].

Today the smectite to illite transition temperature is reached only at a depth of thousands of meters, but on the hot early Earth it must have occurred closer to the surface. The cycle times of both the clay and the adsorbate cycle must have been much shorter due to the higher Rayleigh number of the convection cycle at the higher temperature. In the ocean organic molecules would have been highly diluted, impeding reactions [44-47], but adsorption would have concentrated them. The proposed 'clay-adsorbate soup' consisting of high energy organics desorbed during the transition would differ from the UV-light generated 'primordial soup' proposed by Haldane [10] or the more recently proposed soup for the early ocean [47]. The organisms living today deep in the crust possibly still feed on the clay-adsorbate soup, with their ancestors having followed the descending depth of the transition during Earth's history.

Cycles are of interest for the origin of life [48, 49]. The adsorbate cycle might resemble, or be synchronized, with metabolic cycles such as the reducing tricarboxylic acid cycle, which has been considered as the origin of metabolism [50-53]. Synchronization with a thermal cycle would overcome several objections [54] to the concept of a self-organized primordial metabolic cycle, such as their rarity and the implausibility of some type of reactions in water: partial reactions of the tricarboxylic acid cycle seem plausible during adsorption or hydration-dehydration cycles involving clay minerals.

**5. Biology and the Origin of Life**
Reactions on mineral surfaces, especially on sulfide minerals like pyrite, at hydrothermal vents or mounds on the ocean floor have been invoked in the primordial synthesis of the small biomolecules that constitute the building blocks of large biomolecules such as proteins [44, 46, 49, 53, 55-58]. At temperatures of 100 C the synthesis of amino acids from $H_2$, $CO_2$, $NH_4^+$ and $H_2S$ becomes thermodynamically possible [59].

An initial redox equilibrium on Earth would have been disturbed by the leaking of hydrogen from the atmosphere into outer space [60]. This made the atmosphere more oxidizing than the ocean, and the ocean more oxidizing than the crust. The mentioned vent or mound structures are situated at the ocean-crust interface, where the present redox potential difference may have driven prebiotic chemical reactions. Differences in pH occur as well: due to dissolved $CO_2$ the ocean is moderately acidic, whereas high-temperature vent fluids, on the one hand, are strongly acidic and low-temperature mound fluids, on the other hand, are alkaline [57]. A third driving force is the chemical disequilibrium resulting from cooling. High temperatures break chemical bonds because of the high translation entropy of the resulting fragments. As the temperature is lowered, bond formation can become energetically favorable [59], but overcoming the activation energy of the reaction becomes more difficult. More generally, the requirement of catalysts for all three types of reaction is the motivation for the ongoing investigation of the catalytic capabilities of sulfide minerals [57, 58].

Clay minerals can also function as catalysts [61-63]. Adsorption on clay [64] plays a key role in prebiotic chemistry as well and is also the subject of ongoing research. Both small polypeptides and RNA molecules have been synthesized from derivatized small monomer building blocks [8-9, 65]. The longer the molecule, the more difficult desorption becomes for synthesized biomolecules adsorbed on clay [32]; thermal desorption might solve this problem. Could clay play a role as primordial free energy source as well? Most interestingly, small clay particles can be encapsulated by lipid vesicles (which also can contain RNA) [66]: when the clay particle/lipid vesicle is added to a convection cell, the resulting system may very well function as the SHE pictured in Fig. 6.

      Terrestrial organisms have acquired during evolution a complex metabolic machinery that permits them to react to a wide variety of stimuli, either for information gathering or as energy source. Extraterrestrial organisms may have even more diverse capabilities, since they would have to obtain their energy in very different environments [67]. Switching our attention from distant space to distant time, it is to be expected that the first organisms instead obtained their free energy from simple inanimate self-organizing systems. There are few inanimate sources.

      The energy sources for the origin of life have been reviewed [45]. We start with arguing the implausibility of a role in the origin of life of today's biological energy sources photosynthesis, respiration, and fermentation. The first two depend on the complex, membrane-embedded chemiosmotic machinery, which consists of two components, (*i*) an ATP synthase that transduces the energy of a voltage/pH difference across the membrane into ATP, and (*ii*) one or more proton pumps that generate this voltage/pH difference [68-70].

(1) *Photosynthesis*. In algae and chloroplasts the proton pump is rather complex. Bacterial photosynthesizers are more simple. Even the most simple proton pump that works on light, bacteriorhodopsin [71], needs to be combined with ATP synthase, and therefore involves a composite system that seems far too complex to have emerged from scratch.

(2) *Respiration*. This biological energy source can work on wide ranges of oxidants and reductants [72, 73], including hydrogen [74] and ionizing radiation products [75]. A simple respiration-driven proton pump could consist of a single hydrogenase [76] or an enzyme driven by a redox potential difference such as cytochrome *c* oxidase [77, 78]. Both however require an ATP synthase as partner, and we therefore end up with an implausible complex composite system as well.

(3) *Fermentation*. According to current knowledge, a major role for fermentation as primordial energy source [47] is not reasonable. The reactant is unidentified, would probably be highly diluted, and its high free energy of formation would itself need an explanation. Recently the prebiotic synthesis of pyruvate has however been demonstrated from CO and alkyl thiols, with FeS as catalyst, at high temperature (250 C) and pressure (50-200 MPa) [79]. Pyruvate can yield ATP by fermentation by substrate level phosphorylation, using the enzymes pyruvate:ferredoxin oxidoreductase, hydrogenase and succinyl-CoA synthase, with $CO_2$, $H_2$ and acetate as end products [80]. Other types of fermentation also require only a few enzymes [81]. Nevertheless, even a few enzymes are still too many, and the same argument of the implausibility of the many required components applies to fermentation as an early biological energy source.

      ATP synthesis from ADP and phosphate by ATP synthase involves a condensation reaction. More generally, the synthesis of all the key biomolecules such as proteins and nucleic acids involves condensations of small building blocks [44, 82], reactions that we consider the key step in the origin of life. Such condensations can be effected by physical processes that simultaneously yield the required free energy. Plausible processes are:

(1) *Quenching*. As previously discussed in Footnote 1, during quenching a disequilibrium at low temperature is generated by sudden cooling at high temperature: the high-temperature equilibrium is frozen in. Quenching can occur near geothermal heat sources such as hydrothermal vents. Matsuno *et al*. [83] have demonstrated peptide synthesis during quenching from 200-250 C to 0 C;

(2) *Wet-dry cycling* as occurs in tidal ponds [64, 84]. Key biomolecules such as proteins and nucleic acids are condensation products, the natural result of wet-dry cycling. Chemicals formed during dehydration can have a high free energy when returned to water. This applies to polyphosphates and polypeptides [47, 85, 86]. Polyphosphates still play an extensive role in metabolism, including an energetic one [87], and can associate with nucleic acids [88]. Keefe and Miller [89] suggest however that the prebiotic synthesis of polyphosphates is implausible. Except for the high energy phosphate anhydride bond in ATP, there is little similarity to today's biological energy conversion;

(3) The *mineral-based sorption heat engine* proposed in this study.

     These three processes seem sustainable energy sources for condensation reactions on the early Earth. They may have produced small polypeptides that combined to yield proteins with the capability to generate free energy from the environment. The use of *physical* processes for condensations has the advantage that the required irreversibility is more easily assured. In contrast, during the *chemical* synthesis of dipeptides from amino acids—with (Ni,Fe)S as catalyst and in the presence of CO, with $H_2S$ as condensation agent—the reverse reaction, peptide bond hydrolysis, occurs as well [90]. The same notion may be expressed by the statement "For synthesis reactions that require energy—endergonic reactions—a nonchemical form of energy (e.g., light or shockwaves) has to be converted to chemical energy" [91].

     Cairns-Smith [7] has proposed that during the origin of life the first biological information storage was based on clay minerals, and that later a 'genetic takeover' of this function by nucleic acids took place [7]. Here, we propose an early energy generation by clay, and a similar 'catalytic and energetic takeover' of this energy generation by proteins.

     A role of the SHE in the origin of life is supported by similarities between the SHE4 and a previously proposed ancestor of today's biological energy conversion machinery (Fig. 8), similarities that are in accordance with Orgel's continuity principle during evolution [92]. In the thermosynthesis model for the emergence of bioenergetics, the first enzyme was proposed to have been a heat engine driven by thermal cycling in convecting volcanic hot springs [47, 93-96]. This first enzyme dehydrated bound reactants (or 'substrates' in the terminology of biochemistry), binding them at a low temperature in the enzymatic cleft in which water was absent, performing a dehydration reaction, and releasing the condensation product by a thermal unfolding at a higher temperature.[2]

     A dehydration reaction that would give an early selective advantage is phosphorylation of the molecules of lipid vesicles [66], which stabilizes these vesicles [48, 95].

     The single enzyme was the direct progenitor of the key subunit of the $F_1$ part of the ATP synthase enzyme that operates today according to the binding change mechanism [97] and that plays a key role in the process of chemiosmosis. Chemiosmosis is a major partial process of both photosynthesis and respiration [68-70] that "has ramified into many corners of cellular physiology" [70] but, curiously, is hardly recognized in the field of the origin of life [70] (for exceptions see refs. [45, 56, 57]).

     In the thermosynthesis model the genetic machinery emerged in the RNA World as RNA acquired the ability (by coding for it in a gene, and translating this gene) to synthesize this key subunit of the $F_1$ enzyme [95]. It is noted that redox reactions are absent in the many chemical reactions that support the functions of the genetic machinery such as replication, transcription and translation; instead, the pertinent enzymes often make use of ATP or, to a lesser extent, GTP. The emergence made the acquisition of many proteins/enzymes possible, which allowed in turn the emergence of fermentation and photosynthesis. For the latter a model has been given that starts with a heat engine using the thermotropic lipid phase transition in a

---

[2] The idea of starting life with a single enzyme goes back to Troland's publications in the years 1914-1917 [102-104]. He mentions nucleic acids and the need for an energy source [104]. H.J. Muller objected in 1966 [105] to Troland's lack of distinction between 'gene' and 'enzyme'; in 1966 Cech's discovery of ribozymes was of course still 20 years in the future.

biomembrane [94, 95, 98, 99]. Because of the similarities in its chemiosmotic machinery, respiration could easily have evolved from photosynthesis — and vice versa.

Those organisms that acquired the enzymes to gain free energy from the much energy yielding reactions

$$H_2 + CO_2 \rightarrow CH_4 + H_2O \text{ or } CH_3COOH + H_2O \text{ [46, 59, 73, 100]},$$

must have had a competitive advantage, as $H_2$ [74] and $CO_2$ were widely available and "thermodynamically furthest from equilibrium on the early Earth" [46]. These reactions permit organisms to live deep in the Earth's crust [101]. More generally, near hydrothermal vents the formation of small biomolecules, instead of *costing* free energy, *yields* free energy, a condition that has been compared to a 'free lunch' and that must have considerably simplified running the metabolism of the first organisms [59]. When the metabolic machinery emerged, it may have incorporated in the FeS clusters of ferredoxin the minerals that functioned as catalysts of primordial reactions [46, 49, 55-57]. The little energy yielding reactions that resembled the sorption heat engine may however have left few traces, leaving only the binding change mechanism of today's ATP synthase [97] as relic of this evolutionary scaffold.

## 6. Conclusion

Possibly overlooked because of its simplicity, the sorption heat engine may in fact play a central role in the world. Four kinds can be distinguished. The basic notion is robust and general — additional temperature dependent regulation will be engine-specific and falls outside the scope of the present paper. The engine is already used in refrigeration and chemical separation and purification. Outside engineering, the self-organizing and inanimate engine may be the free energy source that provides solutions for long standing scientific problems such as the origin of life, the first metabolic cycles, the emergence of oil, and the presence of microorganisms in the Earth's crust. A novel unifying point of view is obtained on phenomena in physics, analytical chemistry, chemical and thermal engineering, geology and biology that are commonly considered unrelated.

The thermosynthesis theory proposes that the first enzymes functioned by a thermal variation of the binding change mechanism of today's ATP synthase. These enzymes would have functioned similarly to the proposed sorption heat engines, which may have set the stage for the origin of life.

## Acknowledgements

The authors express their gratitude to Phil Rosenberg for helpful discussions, to Karl Sharrah and Adelina Hristova for proofreading, and to an anonymous reviewer for leading us to additional literature references for sections 4 and 5 and pointing out the need for improving the discussion of primordial redox processes.

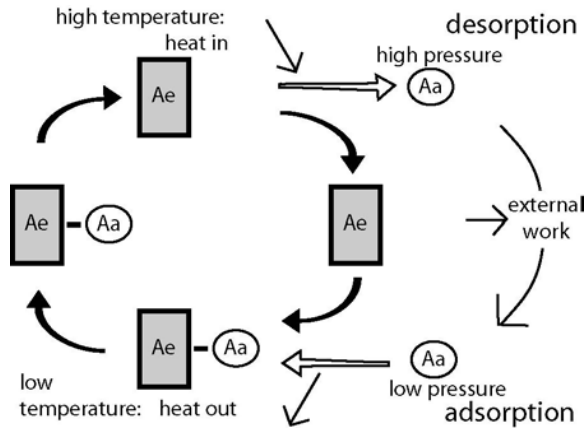

Fig. 1. The sorption heat engine (SHE). The engine works on temperature dependency of adsorption. Heat enters the engine during high-temperature desorption and leaves the engine during low-temperature adsorption. The increased pressure of the desorbed adsorbate (Aa) accounts for the free energy gain. Since both the adsorbent (Ae) and adsorbate remain unchanged during the cycle, the engine is a sorption heat engine of the first kind.

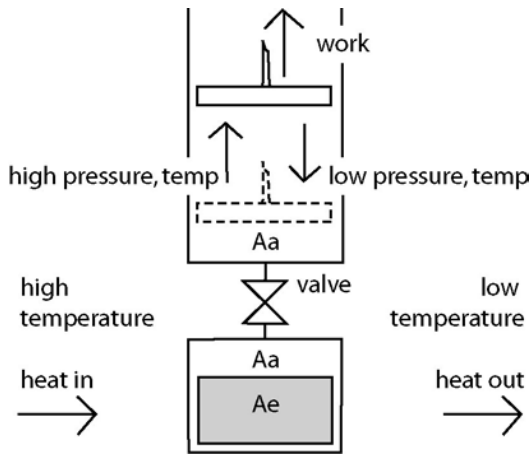

Fig. 2. Working cycle of an SHE of the first kind. With the valve closed, the adsorbent is heated. The resulting desorption of the adsorbate increases the pressure in the lower reservoir. Upon the opening of the valve, the adsorbate enters the upper reservoir, and the increasing pressure moves the piston, which performs external work. Cooling the adsorbent results in adsorption after the opening of the valve, which lowers the pressure. The piston returns to its original position and the valve is closed again.

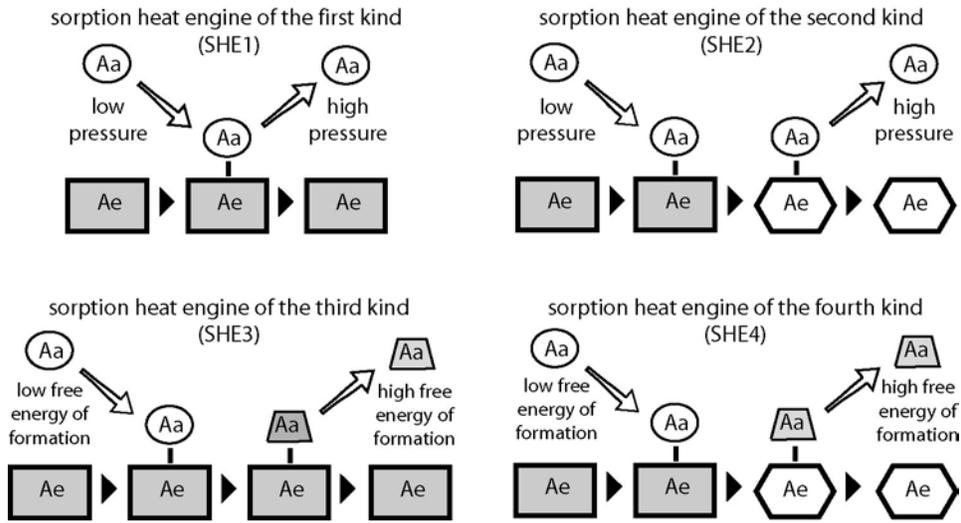

Fig. 3. The four kinds of sorption heat engines. In all engines adsorption occurs at low temperature and desorption at high temperature, and external work is done during the thermal adsorption-desorption cycle, either by an increase in pressure of the adsorbate or by an increase in its free energy of formation. The engines are distinguished by changes—indicated by shape and shading— during the cycle. First kind: no change in adsorbate or adsorbent; second kind: change in adsorbent; third kind: change in adsorbate; fourth kind: change in both adsorbent and adsorbate.

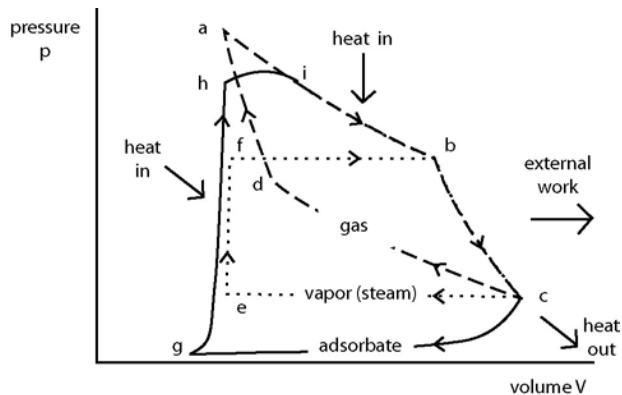

Fig. 4. pressure (*p*)-volume (*V*) diagrams of the working cycles of a Sterling engine, the steam engine and the sorption heat engine, with the respective working fluids gas, vapor and adsorbate. To help the comparison of the 3 graphs, the adiabat b → c, associated with cooling and power generation by expansion, is shared:
A. Gas. This well known graph (a → b → c → d → a) consists of two isotherms and two adiabats;
B. Vapor (steam). f → b: water boils at high, constant *p* and temperature (*T*) while *V* increases and work is done; b: all liquid has turned into steam; b → c: the steam performs external work. Condensation is ignored; c: condensation starts; c → e: because of cooling, steam condenses at low *T* until completion; e → f: water moves to the boiler;
C. Adsorbate (see Figs 1 and 2). g: the valve is closed, at low *T* the adsorbate has been completely adsorbed by the adsorbent; g → h: the adsorbent is heated, Aa desorbs, *p* increases; h: the valve is opened, *V* increases, and work is being done; h → i: additional Aa is desorbed; i: the valve is closed; i → b → c: Aa behaves as a gas; c: cooling of Ae starts, the valve opens; c → g: *p* and *V* decrease. g: the valve is closed.
During the high temperature part of the cycle, the adsorbate behaves as a gas. During the low temperature part, *p* reaches a very low value and the graph approaches the *V* axis.

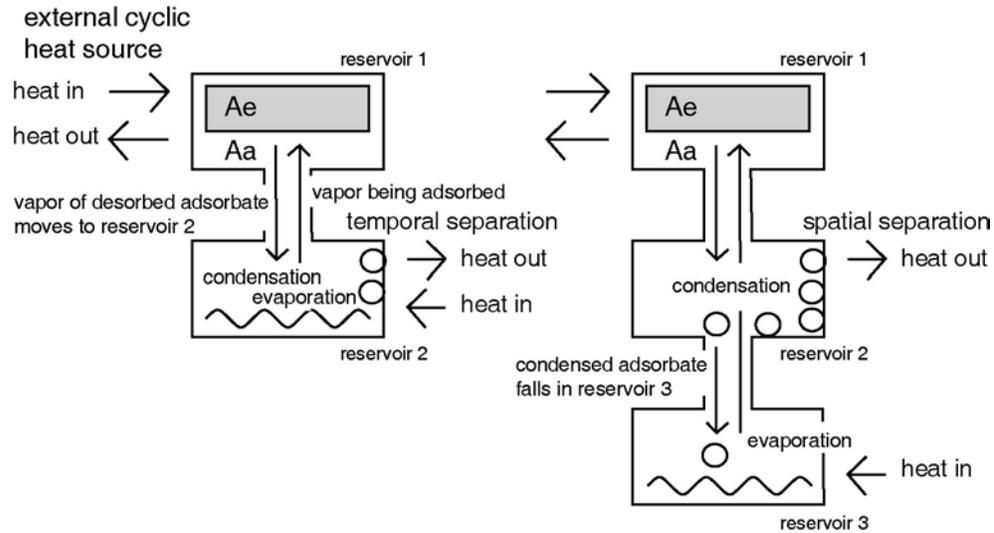

Fig. 5. Simple sorption-based refrigerators (after Fig. 6, ref. [15]). The most simple refrigerator (left) consists of two connected reservoirs containing a vapor as adsorbate. The first reservoir contains the adsorbent and is thermally cycled by an external heat source/sink, which results in cyclic adsorption and desorption of the adsorbate vapor. Desorption increases the pressure of the vapor, causing its condensation in the second reservoir with release of heat. Cooling of the first reservoir causes adsorption, which lowers the vapor pressure again, effecting evaporation in the second reservoir with uptake of heat and thus local cooling. Here, in the left figure, the heat release and uptake are temporally separated; they can be spatially separated (right figure) by adding a third reservoir that collects by gravity the vapor that has condensed in the second reservoir.
The system to the right has found practical application as refrigerator [15], with the first reservoir being heated (by the Sun) and cooled outside the home during the day-night thermal cycle, the second reservoir being placed inside the home, and the third reservoir being thermally connected to the refrigerated space.

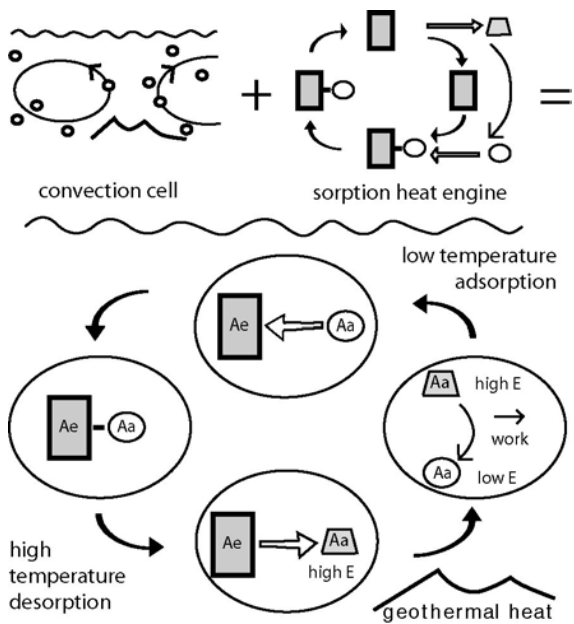

Fig. 6. Merging of the convection and sorption heat engine concepts. The thermal cycling inherent to convection is used to drive the sorption heat engine. The merger of the mechanisms of self-organizing convection and the simple sorption heat engine yields an inanimate engine that may generate the chemical free energy required for the origin of life.

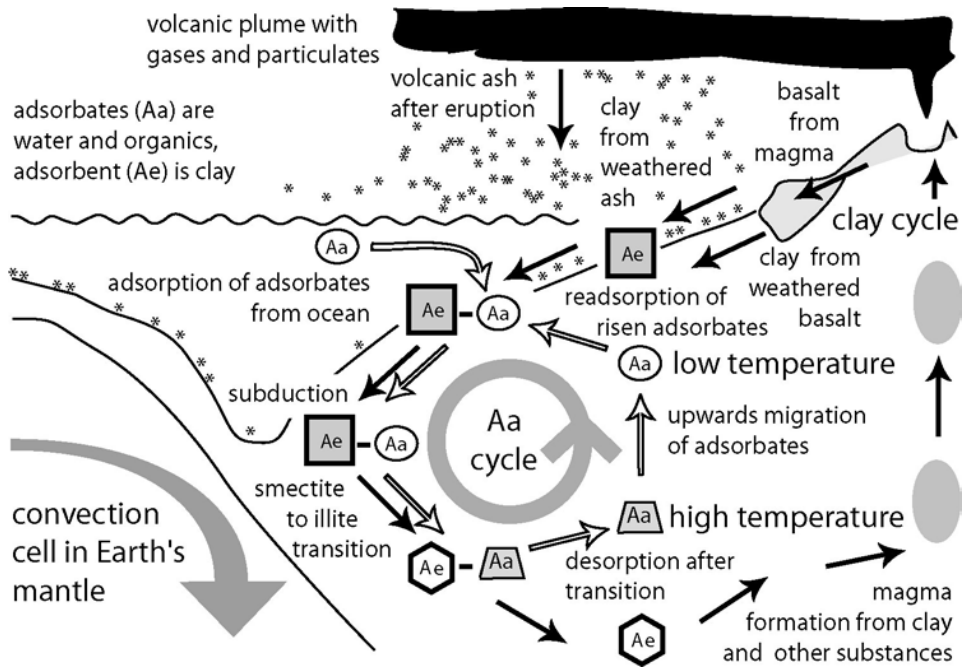

Fig. 7. Proposed adsorbate cycle in the Earth's crust. The causal chain is mantle convection → clay cycle (black arrows) → adsorbate cycle (open arrows). Clay formed from volcanic material (right) is a strong adsorbent of water and organics present in the ocean. Upon heating during subduction, clay undergoes the smectite to illite transition, which diminishes its power of adsorption. The resulting desorbed adsorbates migrate upwards and are readsorbed by cooler clay near the surface, which is in turn subducted. SHE4 activity during the cycle may yield desorbed organics with high chemical free energy on which microorganisms living in the Earth's crust could feed.

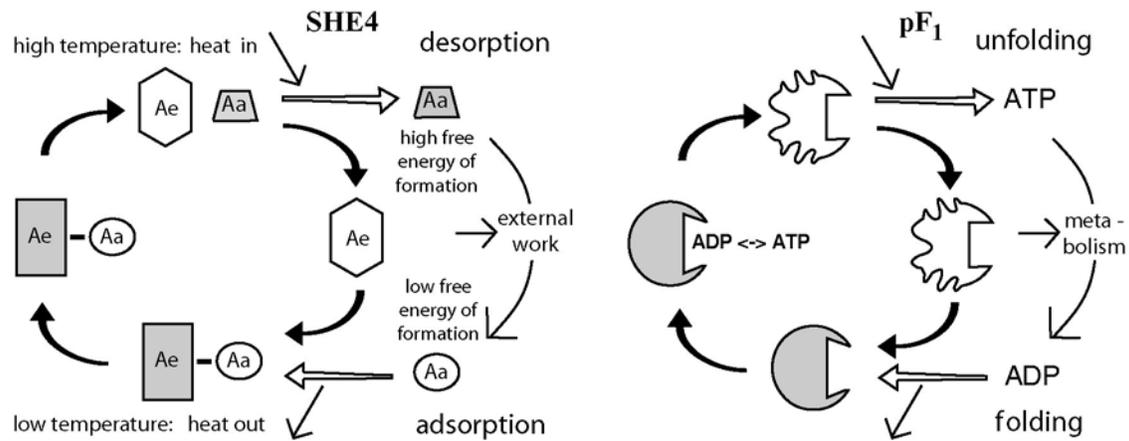

Fig. 8. Similarities between the SHE4, the proposed sorption heat engine of the fourth kind, and pF$_1$, a previously proposed first primordial enzyme [48].  ATP synthesis by the pF$_1$ enzyme was proposed during thermal cycling, the result from it being carried along by convection currents; the enzyme would be a direct ancestor of today's ATP synthase.  Following Boyer's binding change mechanism [97], the enzyme bound ADP and phosphate at low temperature, rejecting heat as it folded.  The strongly bound ADP and phosphate formed strongly bound ATP. At the high temperature the enzyme took up heat, unfolded, and released the high free energy ATP.  The similarities are (1) the use of a thermal cycle; (2) a thermal transition; (3) a heat engine mechanism; (4) temperature-dependent bonding; (5) a product with a high chemical free energy.